\documentclass[10pt,superscriptaddress,twocolumn,amsmath,amssymb,aps,pra]{revtex4}
\usepackage{mathrsfs}
\usepackage{graphicx}
\usepackage{dcolumn}
\usepackage{bm}

\newcommand{\be}{\begin{equation}}
\newcommand{\ee}{\end{equation}}
\newcommand{\bea}{\begin{eqnarray}}
\newcommand{\eea}{\end{eqnarray}}

\begin{document}
\title{Particle and spin transports of spin-orbit coupled Fermi gas through a Quantum Point Contact}
\author{Xiaoyu Dai}
\author{Boyang Liu}\email{boyangleo@gmail.com}
\affiliation{Institute of Theoretical Physics, Faculty of Science, Beijing University of Technology, Beijing, 100124, China}

\date{\today}
\begin{abstract}
The particle and spin transport through a quantum point contact between two Fermi gases with Raman-induced spin-orbit coupling are investigated. We show that the particle and spin conductances both demonstrate the structure of plateau due to the mesoscopic scale of the quantum point contact. Compared with the normal Fermi gases the particle conductance can be significantly enhanced by the spin-orbit coupling effect. Furthermore, the conversion of the particle and spin currents can take place in the spin-orbit coupled system, and we find that it is controlled by the parameter of two-photon detuning. When the parameter of two-photon detuning vanishes the particle and spin currents decouple.
\end{abstract}
 \maketitle
\section{Introduction}
Transport measurements are important tools to investigate the fundamental properties of  states of matter. Recently, the studies of transport in cold atoms have become one of the frontiers in the area. Many experiments have been conducted, including particle transport \cite{Stadler,Krinner1,Valtolina,Husmann,Krinner2,Hausler,Burchianti,Kwon2020,Pace2021}, spin transport \cite{Sommer,Bardon,Koschorreck,Luciuk,Valtolina2017} and heat transport \cite{Brantut,Husmann2018,Hausler2021}. Particularly, the realization of two-terminal set-up by ETH's group \cite{Stadler,Krinner1,Husmann,Krinner2,Hausler,Brantut,Husmann2018} paved a road to extend this paradigmatic tool from condensed matter physics to various unique states in cold atom physics, for instance, the quantized conductance of neutral matter has been observed \cite{Krinner1}, and the anomalous conductance of unitary Fermi gas has been studied both experimentally \cite{Krinner2} and theoretically\cite{Glazman,Liu2017,Uchino2017}.

The interplay between charge and spin degrees of freedom is ubiquitous in physical systems. By utilizing light-atom interactions the synthetic spin-orbit (SO) coupling in neutral atoms has be realized in both bosonic and fermionic cold atom systems \cite{Lin2011,Wang2012,Cheuk2012}. These achievements have stimulated intensive studies in this area \cite{Goldman2014,Zhai2015}. In transport experiments the SO coupling may generate spin and charge currents conversion, which is a key phenomenon in spintronics and can facilitate technological applications, for instance, it's possible to control spin signals by manipulating the electric signals.

In this work we study the spin and particle transports between two Fermi gases with Raman-induced SO coupling. The two reservoirs are connected by a quantum point contact (QPC). We calculate the particle and spin currents using the Keldysh formulism \cite{Kamenev2011}. The chemical potentials of spin-up and spin-down particles are tuned to be different in either of the reservoirs, and then the particle current $I_p\equiv I_\uparrow+I_\downarrow$  and spin current $I_s\equiv I_\uparrow-I_\downarrow$ can be generated. In the linear response regime they can be expressed as
\bea
\left(\begin{array}{c}I_p\\I_s\end{array}\right)=\left(\begin{array}{cc} \sigma_p & \sigma_o \\ \sigma_o^\prime & \sigma_s \end{array}\right)
\left(\begin{array}{c}\Delta \mu_p \\ \Delta \mu_s\end{array}\right),\label{eq:currents}
\eea
where $\Delta\mu_p=\bar\mu_R-\bar\mu_L$ and $\Delta\mu_s=\delta\mu_R-\delta\mu_L$. $\bar\mu_j=(\mu_{j\uparrow}+\mu_{j\downarrow})/2$ and $\delta\mu_j=\mu_{j\downarrow}-\mu_{j\uparrow}$ are the average and difference of the chemical potentials of spin-up and spin-down fermions, where $j=L,R$ denote the left and right reservoir. $\sigma_p$ and $\sigma_s$ are the particle and spin conductances. $\sigma_o$, and $\sigma_o^\prime$ describe the conversion of the particle and spin currents. We investigate the variations of the elements $\sigma_p$, $\sigma_s$, $\sigma_o$, and $\sigma_o^\prime$ with respect to various parameters in system with Raman-induced SO coupling.

\section{Model}
The whole system, including two reservoirs and the QPC, can be described by the following Hamiltonian (setting $\hbar=1$)
\be
\hat{H}=\hat{H}_L+\hat{H}_R+\hat{H}_T,
\label{eq:Ham}
\ee
where $\hat{H}_L (\hat{H}_R)$ describes the Fermi gas with Raman-induced SO coupling in the left (right) reservoir and is given by
\bea
\hat H_j=\Psi^\dagger_j G^{-1}_j \Psi_j,
\eea
where $\Psi^\dagger_j=\left(\begin{array}{cc}\hat\psi^\dagger_{j\uparrow}&\hat\psi^\dagger_{j\downarrow}
\end{array}\right)$ and
\bea
&&G^{-1}_j=\cr&&\left(\begin{array}{cc}
\frac{(k_x+k_0)^2+k^2_\perp}{2m}+\frac{\delta}{2}-\mu_{j\uparrow}&\frac{\Omega}{2}\\ \frac{\Omega}{2} & \frac{(k_x-k_0)^2+k^2_\perp}{2m}-\frac{\delta}{2}-\mu_{j\downarrow}\end{array}\right).
\label{eq:Hreservoir}\cr&&
\eea
The operator $\hat{\psi}^\dagger_{j\sigma}(\hat{\psi}_{j\sigma})$ describes the creation (annihilation) of a fermion atom with spin $\sigma=\uparrow,\downarrow$ in the $j$-th reservoir.  $m$ is the mass of fermions and $k_\perp^2=k_y^2+k_z^2$. $\Omega/2$ is the strength of the Raman coupling and $k_0$ is the wave vector of the laser. $\delta$ is the two-photon detuning \cite{Zhai2015}. With the Pauli matrices the Eq. (\ref{eq:Hreservoir}) can be cast as
\bea
G^{-1}_j=\frac{{\bf k}^2}{2m}+{\bf B}_{\bf k}\cdot {\bf\sigma}+E_r-\bar\mu_j,
\eea
where ${\bf B}_{\bf k}=(\Omega/2, 0,k_xk_0/m+\delta/2-\delta\mu_j/2)$, and $E_r=k_0^2/2m$ is the recoil energy.
With SO coupling the spin is not a good quantum number anymore. However, the single particle Hamiltonian can be diagonalized to two eigenstates with another good quantum number, the helicity `$\pm$', which denote being spin parallel or anti-parallel to the Zeeman field ${\bf B}_{\bf k}$. With a unitary transformation $U_j\hat H_j U^\dagger_j$ the eigenenergies can be obtained as
\bea
E_{{\bf k},\pm}=\frac{{\bf k}^2}{2m}\pm\xi({ k_x})+E_r-\big(\bar\mu_j\mp S({ k_x})\delta\mu_j\big) \label{eq:eigenenergy},
\eea
where
\bea
&&\xi({ k_x})=\sqrt{\Big(k_xk_0/m+\delta/2\Big)^2+\Omega^2/4},\cr
&&S({ k_x})=\frac{k_xk_0/m+\delta/2}{2\sqrt{\Big(k_xk_0/m+\delta/2\Big)^2+\Omega^2/4}}
.\eea
In this work we study the cases where $\delta\mu_j$ is small compared with the chemical potentials $\mu_{j\sigma}$. Hence, in Eq.(\ref{eq:eigenenergy}) and the following work we only keep the terms up to the first order of $\delta\mu_j$.
The fields $\hat\psi_{j\pm}$ corresponding to these two branches can be found by a unitary transformation as the following
\bea
\left(\begin{array}{c}\hat\psi_{j+}({\bf k})\\ \hat\psi_{j-}({\bf k})\end{array}\right) = \left(\begin{array}{cc} D_j({ k_x}) & -O_j({ k_x}) \\  O_j({ k_x}) & D_j({ k_x}) \end{array}\right)
\left(\begin{array}{c}\hat\psi_{j\uparrow}({\bf k})\\ \hat\psi_{j\downarrow}({\bf k})\end{array}\right),\label{eq:unitarytransformation}
\eea and the elements in above matrix are presented as
\begin{align}
D_j({k_x})&=\frac{\Lambda(k_x)}{\sqrt{\Lambda(k_x)^2+\Omega^2/4}}\nonumber\\&+\frac{\Lambda(k_x)\Omega^2/4}{2\xi(k_x)(\Lambda(k_x)^2+\Omega^2/4)^{3/2}}\delta\mu_j,
\nonumber\\ O_j({ k_x})&=\frac{\Omega/2}{\sqrt{\Lambda(k_x)^2+\Omega^2/4}}\nonumber\\&-\frac{\Lambda(k_x)^2\Omega/2}{2\xi(k_x)(\Lambda(k_x)^2+\Omega^2/4)^{3/2}}\delta\mu_j,
\end{align} where
\bea
\Lambda(k_x)=k_xk_0/m+\delta/2+\xi(k_x).
\eea

The tunneling between two reservoirs through the QPC is described by $\hat{H}_T$. In real space it can be written as
\bea
\hat H_T=&&\sum_{n=0}^\infty[\mathcal T_n^{(+)}\hat\psi_{L+}^\dagger(0)\hat \psi_{R+}(0)
+\mathcal T_n^{(-)}\hat\psi_{L-}^\dagger(0)\hat \psi_{R-}(0)]\cr&&+h.c..\label{eq:tunneling}
\eea Here we assume the eigenstates $\psi_{R+}$ and $\psi_{R-}$ are transported through point ${\bf x}=0$ between the two reservoirs. In the experimental setup the QPC is formed by the confinement in $\hat y$ and $\hat z$ directions, which lead to the transport channels with energies of  $\epsilon_\perp(n_y,n_z)\equiv (\frac{1}{2}+n_y)\omega_y+(\frac{1}{2}+n_z)\omega_z+V_g$ \cite{Krinner1,Krinner2}. For simplicity an effective gate potential $\bar V_g=V_g+\frac{1}{2} \omega_y+\frac{1}{2} \omega_z$ can be defined and we assume $\omega_z\gg\omega_y$, then the several lowest transport channels would be $n_y \omega_y+\bar V_g$, and they are non-degenerate. Then the tunneling amplitude $\mathcal T_n^{(\pm)}$ can be written as
\bea
\mathcal T_n^{(\pm)}({\bf k}_L,{\bf k}_R)=\mathcal{T} \prod_{j=L,R}\Theta(\epsilon_{\pm}({\bf k}_j)-n \omega_y-\bar V_g),
\eea
where $\epsilon_{\pm}({\bf k})=\frac{{\bf k}^2}{2m}\pm\xi(k_x)+E_r$ is the single particle energy of field $\hat\psi_{j\pm}$ and $\Theta(\epsilon_{\pm}({\bf k}_j)-n \omega_y-\bar V_g)$ is the heaviside step function. Above tunneling amplitude indicates that only the particle with energy $\epsilon_{{\bf k},\pm}>n \omega+\bar V_g$ can enter the $n$-th cannel and will come out from the same channel. That is, there is no inter-channel scattering within the QPC region.

The current for spin-$\sigma$ is defined as
\bea
I_\sigma\equiv\frac{1}{2}\left\langle\frac{\partial}{\partial t}(N_{L\sigma}-N_{R\sigma})\right\rangle,
\eea
where $N_{j\sigma}\equiv \sum_{{\bf k}\sigma}\hat{\psi}^\dag_{j\sigma}({\bf k})\hat{\psi}_{j\sigma}({\bf k})$. In above expression the averages $\langle\cdot\cdot\cdot\rangle$ is taken over a time-evolving many-body state, which can be calculated using the Keldysh formalism. Please see the appendix A for details. In the linear response regime the particle and spin currents are expressed as the Eq. (\ref{eq:currents}). The elements $\sigma_p$, $\sigma_s$, $\sigma_o$, and $\sigma_o^\prime$ in Eq. (\ref{eq:currents}) can be calculated as the following
\begin{align}
\sigma_p=&\frac{\alpha}{h}\sum_{n=0}^\infty\Bigg\{\int_{\epsilon_+(k_{1x})<\epsilon_F}dk_{1x}\int^{\infty}_{-\infty}dk_{2x}\nonumber\\&\Theta(\epsilon_+(k_{2x})-n\omega_y-\bar V_g)F_1(k_{1x},k_{2x})n_+(k_{2x})\nonumber\\&
+\int_{\epsilon_-(k_{1x})<\epsilon_F}dk_{1x}\int^{\infty}_{-\infty}dk_{2x}\nonumber\\&\Theta(\epsilon_-(k_{2x})-n\omega_y-\bar V_g)F_1(k_{1x},k_{2x})n_-(k_{2x})\Bigg\},
\end{align}
\begin{align}
\sigma_o=&\frac{\alpha}{h}\sum_{n=0}^\infty\Bigg\{-\int_{\epsilon_+(k_{1x})<\epsilon_F}dk_{1x}\int^{\infty}_{-\infty}dk_{2x}\nonumber\\&\Theta(\epsilon_+(k_{2x})-n\omega_y-\bar V_g)F_1(k_{1x},k_{2x})S(k_{2x})n_+(k_{2x})\nonumber\\&
+\int_{\epsilon_-(k_{1x})<\epsilon_F}dk_{1x}\int^{\infty}_{-\infty}dk_{2x}\nonumber\\&\Theta(\epsilon_-(k_{2x})-n\omega_y-\bar V_g)F_1(k_{1x},k_{2x})S(k_{2x})n_-(k_{2x})\Bigg\},
\end{align}
\begin{align}
\sigma^\prime_o=&\frac{\alpha}{h}\sum_{n=0}^\infty\Bigg\{\int_{\epsilon_+(k_{1x})<\epsilon_F}dk_{1x}\int^{\infty}_{-\infty}dk_{2x}\nonumber\\&\Theta(\epsilon_+(k_{2x})-n\omega_y-\bar V_g)F_2(k_{1x},k_{2x})n_+(k_{2x})\nonumber\\&
+\int_{\epsilon_-(k_{1x})<\epsilon_F}dk_{1x}\int^{\infty}_{-\infty}dk_{2x}\nonumber\\&\Theta(\epsilon_-(k_{2x})-n\omega_y-\bar V_g)F_2(k_{1x},k_{2x})n_-(k_{2x})\Bigg\},
\end{align}
\begin{align}
\sigma_s=&\frac{\alpha}{h}\sum_{n=0}^\infty\Bigg\{-\int_{\epsilon_+(k_{1x})<\epsilon_F}dk_{1x}\int^{\infty}_{-\infty}dk_{2x}\nonumber\\&\Theta(\epsilon_+(k_{2x})-n\omega_y-\bar V_g)F_2(k_{1x},k_{2x})S(k_{2x})n_+(k_{2x})\nonumber\\&
+\int_{\epsilon_-(k_{1x})<\epsilon_F}dk_{1x}\int^{\infty}_{-\infty}dk_{2x}\nonumber\\&\Theta(\epsilon_-(k_{2x})-n\omega_y-\bar V_g)F_2(k_{1x},k_{2x})S(k_{2x})n_-(k_{2x})\Bigg\},
\end{align}
where
\bea
&&n_\pm(k_x)=\frac{1}{\exp\{(\epsilon_\pm(k_x)-\epsilon_F)/k_BT\}+1}\cr&&
F_1(k_{1x},k_{2x})=\frac{(\Lambda(k_{1x})\Lambda(k_{2x})+\Omega^2/4)^2}{2(\Lambda(k_{1x})^2+\Omega^2/4)(\Lambda(k_{2x})^2+\Omega^2/4)},\cr&&
F_2(k_{1x},k_{2x})=\frac{\Lambda(k_{1x})^2\Lambda(k_{2x})^2-\Omega^4/16}{2(\Lambda(k_{1x})^2+\Omega^2/4)(\Lambda(k_{2x})^2+\Omega^2/4)}.\cr&&
\eea
In this work we use $\bar \mu_L$ as the energy scale and define the Fermi energy as $\epsilon_F=\bar\mu_L$. The so-called transparency is defined as $\alpha=\frac{|\mathcal T|^2m^3\epsilon_F}{\pi^2}$. Here we assume perfectly transparent junction and set $\alpha=1$.

\section{Particle and spin conductances for $\delta=0$}
In this section we investigate the variation of the conductance matrix in Eq. (\ref{eq:currents}) for the case of symmetric dispersion, where the two-photo detuning $\delta=0$. Several properties of the conductances can be found.

\begin{figure}[h]
\includegraphics[width=0.48\textwidth]{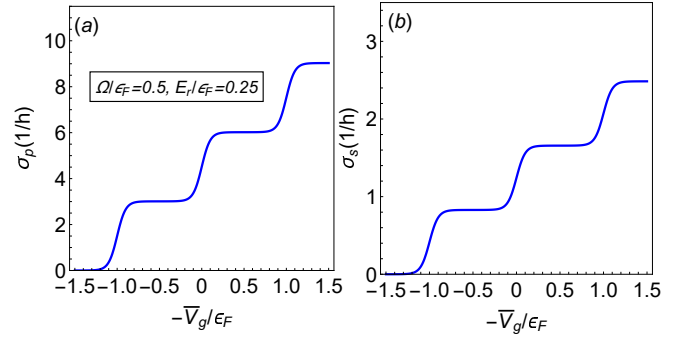}
\caption{(Color online) The particle conductance $\sigma_p$ and spin conductance $\sigma_s$ as functions of gate potential $\bar V_g$ for fixed $\Omega$ and $E_r$. }
\label{fig:plateau}
\end{figure}
First, the particle conductance $\sigma_p$ demonstrates the structure of quantized plateaus analogous to the system without SO coupling \cite{Krinner1,Landauer,Buttiker}, except that the height of the plateau is not $2/h$ any more. For instance, the height of the plateau can be larger than $2/h$ for $\Omega/\epsilon_F=0.5$ and $E_r/\epsilon_F=0.25$ as shown in Fig.\ref{fig:plateau} (a). Furthermore, the spin conductance also shows plateau structure as shown in Fig.\ref{fig:plateau} (b).

\begin{figure}[b]
\includegraphics[width=0.48\textwidth]{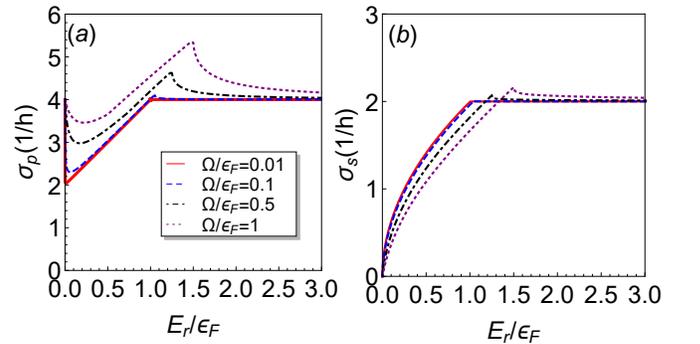}
\caption{(Color online) The particle conductance $\sigma_p$ and spin conductance $\sigma_s$ as functions of $E_r$ for $\Omega/\epsilon_F=0.01$, $0.1$, $0.5$ and $1$. In both (a) and (b), we set $\bar V_g/\epsilon_F=0.5$ and $\delta=0$. }
\label{fig:cond0}
\end{figure}
\begin{figure}[t]
\includegraphics[width=0.35\textwidth]{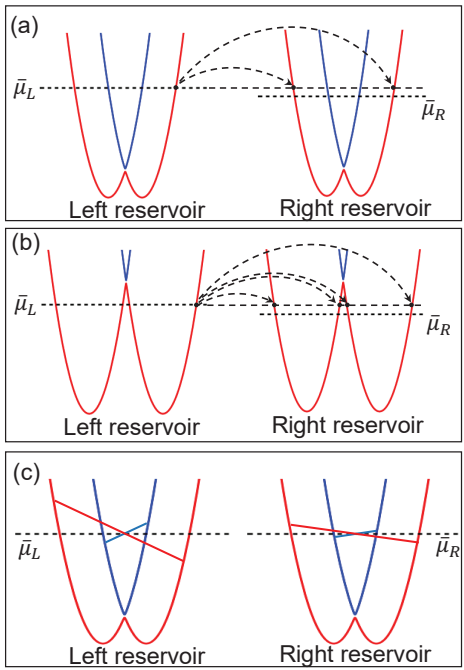}
\caption{(Color online) the schematic plot of dispersions and Fermi surfaces for the region of (a) $\Omega\ll E_r\ll\epsilon_F$ and (b) $\Omega\ll\epsilon_F\ll E_r$, the red part of the dispersion represents the lower branch and the blue one represents the upper branch. For the region (a) a particle around the Fermi surface on the lower branch can be transported to two states in the right reservoir. For the region (b) a particle on the lower branch can be transported to four states in the right reservoir. In both (a) and (b) we set $\delta\mu_j=0$. (c), the schematic plot of the tilted Fermi surface due to the non-zero $\delta\mu_j$. The red straight line represents the Fermi surface of the lower branch, and the blue one is for the upper branch.  }
\label{fig:sketch}
\end{figure}
Second, fixing $\bar V_g/\epsilon_F=0.5$ the particle and spin conductances are at the first plateau, then we can study the variation of the height of the plateau by changing the parameters. In Fig. \ref{fig:cond0} (a) and (b) we plot $\sigma_p$ and $\sigma_s$ as functions of $\Omega$ and $E_r$. For the case of small $\Omega$, for instance, the curve of $\Omega/\epsilon_F=0.01$ in  Fig. \ref{fig:cond0} (a), one observes that $\sigma_p$ can reach $4/h$ when $E_r$ approaches 0, which is double of the height of the plateau in the system without SO coupling \cite{Krinner1,Landauer,Buttiker}. That is, in the region of $E_r\ll\Omega$ the coupling of spin-up and spin-down fields can enhance the particle transport ability. Then, when $E_r$ increases to the region $\Omega\ll E_r\ll\epsilon_F$ the particle conductance $\sigma_p$ drops rapidly and can reach the value of $2/h$. In this region the spin-up and spin-down fields are roughly decoupled. It's reduced to a system without SO coupling. Hence, $\sigma_p$ reaches the value of $2/h$, which is the conductance of normal gas.  When $E_r$ increases further and becomes comparable to $\epsilon_F$ one observes that $\sigma_p$ increases up to $4/h$ again. This can be explained by the graphs of Fig. \ref{fig:sketch} (a) and (b), in which we plot the transport of particles around the Fermi surface in one-dimensional case. The single particle dispersion of region $E_r\ll\epsilon_F$ is sketched in Fig. \ref{fig:sketch} (a). One observes that a particle on the lower branch in the left reservoir can be transported to two states in the right reservoir, since the momentum is not conserved in the tunneling process as show in Eq. (\ref{eq:tunnelingmomentum}) in the appendix A. For the region $E_r\gg\epsilon_F$  in Fig. \ref{fig:sketch} (b), we demonstrate that a particle on the lower branch in the left reservoir can be transported to four state in the right reservoir. This explains the increasing of the conductance as $E_r$ increases. Fig. \ref{fig:sketch} (a) and (b) are examples in one-dimensional case. Our system is three dimensional. The Fermi surface is not a set of points but a sphere. The situation is more complicated but the logic beneath is the same. Comparing the four curves of $\Omega/\epsilon_F=0.01,0.1,0.5,$ and $1$ in Fig. \ref{fig:cond0} (a) one observe that for any fixed $E_r$ the particle conductance $\sigma_p$ increases as $\Omega$ increases. There are two effects that help to increase the particle conductance. One is the the mixing of spin-up and spin-down fields. The other effect is the density of state. The low energy density of state in Raman-induced SO coupled system is larger than the case of $k_x^2$ dispersion \cite{Zhai2015}. The curves of $\Omega/\epsilon_F=0.5$ and $1$ demonstrated clear peaks, which correspond to the cases when the Fermi surfaces reach the middle peak of the dispersion of the lower branch, where the density of state is large.

Third, in Fig. \ref{fig:cond0} (b) one observes that the spin conductance $\sigma_s$ increases as $E_r$ increases and saturates at the value of $2/h$. The variation of $\Omega$ doesn't affect $\sigma_s$ much.

Forth, our calculation shows that the off-diagonal terms $\sigma_o$ and $\sigma_o^\prime$ both vanish in the case of $\delta=0$. The spin current is driven by the bias of the spin difference $\Delta\mu_s=\delta\mu_R-\delta\mu_L$. In Eq. (\ref{eq:eigenenergy}) we can see that the spin difference $\delta\mu_j$ deforms the Fermi surface. Roughly speaking, it tilts the Fermi surface as sketched in Fig. \ref{fig:sketch}(c). During the spin transport, the Fermi surface of the left reservoir will become more horizontal. The particles on the lower branch with momenta $k_x<0$ in the left reservoir will be transported to the right, while the particles with momenta $k_x>0$ in the right reservoir will be transported to the left. Since the dispersion is symmetric about the axis $k_x=0$, the number of particles transported from the left to the right reservoir equals to the one transported from the right to the left reservoir. Hence, the bias of the spin difference $\Delta\mu_s$ will not drive a net particle current. Analogously, $\Delta\mu_p$ will not drive a spin current. Hence, $\sigma_o$ and $\sigma_o^\prime$ are both zero for $\delta=0$ case.

\section{Particle and spin conductances for $\delta\neq0$}
\begin{figure}[t]
\includegraphics[width=0.48\textwidth]{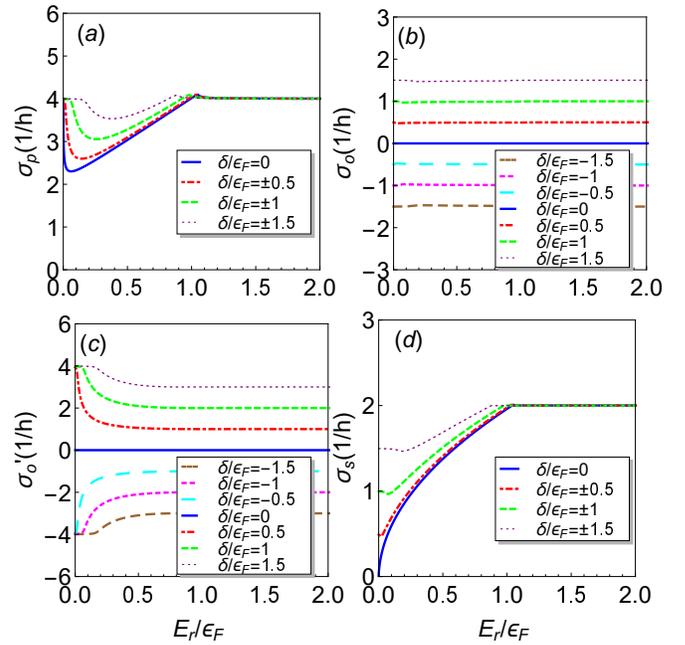}
\caption{(Color online) $\sigma_p$, $\sigma_o$, $\sigma_o^\prime$, and $\sigma_s$ as functions of $E_r$ for different values of $\delta$. Here we set $\Omega/\epsilon_F=0.1$. When $\delta$ changes sign, $\sigma_p$ and $\sigma_s$ remain the same, while $\sigma_o$ and $\sigma_o^\prime$ changes sign.  }
\label{fig:condN0}
\end{figure}
As we discussed in the last section the particle and spin currents decouple when $\delta$ vanishes. In this section we investigate how a non-zero $\delta$ affect the transport. In Fig. \ref{fig:condN0} we plot $\sigma_p$, $\sigma_o$, $\sigma_o^\prime$, and $\sigma_s$ as functions of $E_r$ and $\delta$ for a typical value of $\Omega/\epsilon_F=0.1$. One observe that as $\delta$ increases the $\sigma_p$ and $\sigma_s$ are both enhanced in Fig. \ref{fig:condN0} (a) and (d). Of particular interests are the graph Fig. \ref{fig:condN0} (b) and (c), in which we see that $\sigma_o$ and $\sigma_o^\prime$ become non-zero for non-zero $\delta$, that is, the strength of the coupling between particle and spin currents can be controlled by the two photon detuning $\delta$. Furthermore, the signs of $\sigma_o$ and $\sigma_o^\prime$ are also tunable. When $\delta$ changes sign, $\sigma_o$ and $\sigma_o^\prime$ change signs either, which gives more flexibility to manipulate the conversion of the particle and spin currents.

\section{Conclusions}
In summary, the particle and spin transports through a QPC have been studied in a fermioninc system with Raman-induced SO coupling. Due to the
mesoscopic scale of the QPC both particle and spin conductance exhibit plateau structure. Compared with the system without SO couplings the height of the plateau of the particle conductance $\sigma_p$ is enlarged by the SO coupling effects. The magnitude of the spin conductance $\sigma_s$ majorly depends on the laser wave vector $k_0$. As $k_0$ increases $\sigma_s$ increases and saturates at a fixed value. Large two-photon detuning $\delta$ can help to enhance $\sigma_s$ when $k_0$ is small.  Furthermore, the conversion of the particle and spin currents are exclusively controlled by the two-photon detuning $\delta$. When $\delta$ is non-zero, the particle and spin currents couple together.

\section{Acknowledgements}
The work is supported by the National Science Foundation of China (Grant No. NSFC-11874002), Beijing Natural Science Foundation (Grand No. Z180007).

\appendix
\section{The calculation of current in Keldysh formulism}
To calculate the particle and spin currents we employ the Keldysh formulism in which the whole system is on a closed time contour. The partition function can be written as
\begin{align} \mathcal Z=\frac{1}{\mathcal Z_0}\int D[\bar\psi_{j+},\psi_{j+}, \bar\psi_{j-},\psi_{j-}]\exp(iS),
\end{align}
where $S=S_0+S_T+S_s$ and $S_0$, $S_T$ and $S_s$ correspond to the free,  tunneling and source terms, respectively. $S_0$ can be expressed in the momentum space as the following
\bea
&&S_0 = \cr&&\int d{\bf k} d\omega\sum_{j=L,R}\Big\{\bar\Psi_{j+} [G_{j+}]^{-1}\Psi_{j+}+\bar\Psi_{j-} [G_{j-}]^{-1}\Psi_{j-}\Big\},\cr&&\cr&&
\eea where the fields are given by
\bea
&&\bar\Psi_{j+}=\left(\begin{array}{cc}\bar\psi_{j1+}&\bar\psi_{j2+}\end{array}\right),
\Psi_{j+} =\left(\begin{array}{c}\psi_{j1+}\\ \psi_{j2+}\end{array}\right)\cr&&\bar\Psi_{j-}=\left(\begin{array}{cc}\bar\psi_{j1-}&\bar\psi_{j2-}\end{array}\right),
\Psi_{j-} =\left(\begin{array}{c}\psi_{j1-}\\ \psi_{j2-}\end{array}\right).
\eea
The field $\psi_{j1\pm}$ and  $\psi_{j2\pm}$ are the Keldysh rotation of the fields $\psi_{j\pm}^+$ and $\psi_{j\pm}^-$, which are the fields of $\psi_{j\pm}$ on forward and backward time directions. The Keldysh rotation are given by
\bea
&&\psi_{j1\pm}=\frac{1}{\sqrt 2}(\psi^+_{j\pm}+\psi^-_{j\pm}),~~\psi_{j2\pm}=\frac{1}{\sqrt 2}(\psi^+_{j\pm}-\psi^-_{j\pm}),\cr&&
\bar\psi_{j1\pm}= \frac{1}{\sqrt 2}(\bar\psi^+_{j\pm}-\bar\psi^-_{j\pm}),~~\bar\psi_{j2\pm}=\frac{1}{\sqrt 2}(\bar\psi^+_{j\pm}+\bar\psi^-_{j\pm}).\cr&&
\eea
The Green's functions in the Keldysh space are expressed as
\bea
G_{j\pm} =\left(\begin{array}{cc}G^R_{j\pm}&G^K_{j\pm}\\0&G^A_{j\pm}\end{array}\right),
\eea
The retard(advanced) Green's function is given by
\bea
&&G_{j\pm}^R=\cr&&\frac{1}{\omega-{\bf k}^2/2m\mp\xi(k_x)-E_r+(\bar\mu_j\mp S(k_x))\delta\mu_j)+i0^+},\cr&&
G_{j\pm}^A=\cr&&\frac{1}{\omega-{\bf k}^2/2m\mp\xi(k_x)-E_r+(\bar\mu_j\mp S(k_x))\delta\mu_j)-i0^+}\cr&&
\eea and the Keldysh Green function is
\bea
G^K_{j\pm}=(1-2n_F(\omega))(G^R_{j\pm}-G^A_{j\pm}).
\eea

To write down the action $S_T$ for the tunneling Hamiltonian $\hat H_T$ we use the single particle Hamiltonian $\hat H+\sum_j(\mu_{j\uparrow}\hat N_{j\uparrow}+\mu_{j\downarrow}\hat N_{j\downarrow})$ to construct the time evolution operator
\bea
U(t)=e^{i[\hat H+\sum_j(\mu_{j\uparrow}\hat N_{j\uparrow}+\mu_{j\downarrow}\hat N_{j\downarrow})]t}
\eea Then the time evolution of the tunneling part is given by $H_T(t)=U(t)H_TU^{\dagger}(t)$. In momentum space it's written as
\bea
&&\hat H_T=\sum_{n=0}^\infty\int \frac{d\omega}{2\pi} \frac{d^3{\bf k}_L}{(2\pi)^3}  \frac{d^3{\bf k}_R}{(2\pi)^3}\cr &&[\mathcal T_n^{(+)}\hat\psi_{L+}^\dagger(\omega,{\bf k}_L)\hat \psi_{R+}(\omega-\Delta\mu_p+S(k_{Rx})\Delta\mu_s,{\bf k}_R)\cr&&
+\mathcal T_n^{(-)}\hat\psi_{L-}^\dagger(\omega,{\bf k}_L)\hat \psi_{R-}(\omega-\Delta\mu_p-S(k_{Rx})\Delta\mu_s,{\bf k}_R)]\cr&&+h.c.\cr&& \label{eq:tunnelingmomentum}
\eea

The particle and spin currents are defined in Eq. (\ref{eq:currents}). In momentum space they can be expressed in terms of field $\hat \psi_{j\pm}$ as the following
\begin{widetext}
\bea
I_{\uparrow}(\Omega)=&&-i \sum_{n=0}^\infty\int \frac{d\omega}{2\pi} \frac{d^3{\bf k}_L}{(2\pi)^3}  \frac{d^3{\bf k}_R}{(2\pi)^3}\cr &&\Big\{\big(D_LD_R+O_LO_R\big)\Big(D_LD_R\mathcal T_n^{(+)}\langle\hat\psi_{L+}^\dagger(\omega,{\bf k}_L)\hat \psi_{R+}(\Omega+\omega-\Delta\mu_p+S(k_{Rx})\Delta\mu_s,{\bf k}_R)\rangle
\cr&&+O_LO_R\mathcal T_n^{(-)}\langle\hat\psi_{L-}^\dagger(\omega,{\bf k}_L)\hat \psi_{R-}(\Omega+\omega-\Delta\mu_p-S(k_{Rx})\Delta\mu_s,{\bf k}_R)\rangle\Big)\Big\}+h.c.
,
\cr I_{\downarrow}(\Omega)=&&-i \sum_{n=0}^\infty\int \frac{d\omega}{2\pi} \frac{d^3{\bf k}_L}{(2\pi)^3}  \frac{d^3{\bf k}_R}{(2\pi)^3}\cr &&\Big\{\big(D_LD_R+O_LO_R\big)\Big(O_LO_R\mathcal T_n^{(+)}\langle\hat\psi_{L+}^\dagger(\omega,{\bf k}_L)\hat \psi_{R+}(\Omega+\omega-\Delta\mu_p+S(k_{Rx})\Delta\mu_s,{\bf k}_R)\rangle
\cr&&+D_LD_R\mathcal T_n^{(-)}\langle\hat\psi_{L-}^\dagger(\omega,{\bf k}_L)\hat \psi_{R-}(\Omega+\omega-\Delta\mu_p-S(k_{Rx})\Delta\mu_s,{\bf k}_R)\rangle\Big)\Big\}+h.c.
\eea
\end{widetext}
In above equations the $D_j$ and $O_j$ are the elements of the unitary matrix in Eq.(\ref{eq:unitarytransformation}). Then in the Keldysh formulism the action parts of the tunneling term and source term can be cast as
\begin{widetext}
\bea
&&S_T=\sum_{n=0}^\infty\int \frac{d\omega}{2\pi} \frac{d^3{\bf k}_L}{(2\pi)^3}  \frac{d^3{\bf k}_R}{(2\pi)^3} \Big(J^q_+(0,\omega,{\bf k}_L,{\bf k}_R)+J^q_-(0,\omega,{\bf k}_L,{\bf k}_R)\Big)+h.c.,\cr&&
S_s=-i\sum_{n=0}^\infty\int \frac{d\Omega}{2\pi}\frac{d\omega}{2\pi} \frac{d^3{\bf k}_L}{(2\pi)^3}  \frac{d^3{\bf k}_R}{(2\pi)^3}\Big\{\cr&&~~~~~~~~~~A^{cl}_\uparrow(\Omega)[(D_LD_R+O_LO_R)(D_LD_RJ^{q}_+(\Omega,\omega,{\bf k}_L,{\bf k}_R)+O_LO_RJ^{q}_-(\Omega,\omega,{\bf k}_L,{\bf k}_R))]\cr&&~~~~~~~+A^{q}_\uparrow(\Omega)[(D_LD_R+O_LO_R)(D_LD_RJ^{cl}_+(\Omega,\omega,{\bf k}_L,{\bf k}_R)+O_LO_RJ^{cl}_-(\Omega,\omega,{\bf k}_L,{\bf k}_R))]\cr&&~~~~~~~+A^{cl}_\downarrow(\Omega)[(D_LD_R+O_LO_R)(O_LO_RJ^{q}_+(\Omega,\omega,{\bf k}_L,{\bf k}_R)+D_LD_RJ^{q}_-(\Omega,\omega,{\bf k}_L,{\bf k}_R))]\cr&&~~~~~~~+A^{q}_\downarrow(\Omega)[(D_LD_R+O_LO_R)(O_LO_RJ^{cl}_+(\Omega,\omega,{\bf k}_L,{\bf k}_R)+D_LD_RJ^{cl}_-(\Omega,\omega,{\bf k}_L,{\bf k}_R))]\Big\}
+h.c.,\eea
where
\bea
&&J^{cl}_+(\Omega,\omega,{\bf k}_L,{\bf k}_R)=\sum_{n=0}^\infty \mathcal T_n^{(+)}\cr&&~~~~~ \Big(\bar\psi_{L1+}(\omega,{\bf k}_L) \psi_{R2+}(\Omega+\omega-\Delta\mu_p+S(k_{Rx})\Delta\mu_s,{\bf k}_R)+\bar\psi_{L2+}(\omega,{\bf k}_L) \psi_{R1+}(\Omega+\omega-\Delta\mu_p+S(k_{Rx})\Delta\mu_s,{\bf k}_R)\Big),
\cr&&J^{cl}_-(\Omega,\omega,{\bf k}_L,{\bf k}_R)=\sum_{n=0}^\infty \mathcal T_n^{(-)}\cr&&~~~~~\Big(\bar\psi_{L1-}(\omega,{\bf k}_L) \psi_{R2-}(\Omega+\omega-\Delta\mu_p-S(k_{Rx})\Delta\mu_s,{\bf k}_R)+\bar\psi_{L2-}(\omega,{\bf k}_L) \psi_{R1-}(\Omega+\omega-\Delta\mu_p-S(k_{Rx})\Delta\mu_s,{\bf k}_R)\Big),\cr&&
J^{q}_+(\Omega,\omega,{\bf k}_L,{\bf k}_R)=\sum_{n=0}^\infty \mathcal T_n^{(+)}\cr&&~~~~~\Big(\bar\psi_{L1+}(\omega,{\bf k}_L) \psi_{R1+}(\Omega+\omega-\Delta\mu_p+S(k_{Rx})\Delta\mu_s,{\bf k}_R)+\bar\psi_{L2+}(\omega,{\bf k}_L) \psi_{R2+}(\Omega+\omega-\Delta\mu_p+S(k_{Rx})\Delta\mu_s,{\bf k}_R)\Big),
\cr&&J^{q}_-(\Omega,\omega,{\bf k}_L,{\bf k}_R)=\sum_{n=0}^\infty \mathcal T_n^{(-)}\cr&&~~~~~\Big(\bar\psi_{L1-}(\omega,{\bf k}_L) \psi_{R1-}(\Omega+\omega-\Delta\mu_p-S(k_{Rx})\Delta\mu_s,{\bf k}_R)+\bar\psi_{L2-}(\omega,{\bf k}_L) \psi_{R2-}(\Omega+\omega-\Delta\mu_p-S(k_{Rx})\Delta\mu_s,{\bf k}_R)\Big).\cr&&
\eea
\end{widetext}
The currents of spin-up and spin-down can be calculated by
\bea
I_\sigma(t)=\int \frac{\Omega}{2\pi}e^{i\Omega t}I_\sigma(\Omega),
\eea
where $I_\sigma(\Omega)$ is given by
\bea
I_\sigma(\Omega)=\left.\frac{i}{2}\frac{\partial \mathcal Z}{\partial A^q_\sigma}\right|_{A_{\uparrow}^{q},A_{\downarrow}^{q},A_{\uparrow}^{cl},A_{\downarrow}^{cl}=0}.
\eea

\end{document}